\documentclass[vecphys]{svmult}

\usepackage{makeidx}         % allows index generation
\usepackage{graphicx}        % standard LaTeX graphics tool
\usepackage{multicol}        % used for the two-column index
\usepackage[bottom]{footmisc}% places footnotes at page bottom
\makeindex             % used for the subject index
\begin{document}

\title*{Interactions and star formation activity \\ in Wolf-Rayet galaxies}
% Use \titlerunning{Short Title} for an abbreviated version of
% your contribution title if the original one is too long
\author{\'Angel R. L\'opez-S\'anchez\inst{1} \and C\'esar Esteban\inst{1}}
% Use \authorrunning{Short Title} for an abbreviated version of
% your contribution title if the original one is too long
\institute{Instituto de Astrof\'{\i}sica de Canarias (IAC), C. V\'{\i}a L\'actea, S/N, 38205 \\ La Laguna, Sta. Cruz de Tenerife, SPAIN.
\texttt{angelrls@iac.es, cel@iac.es}
%\and Name and Address of your Institute \texttt{name@email.address}
}
%
% Use the package "url.sty" to avoid
% problems with special characters
% used in your e-mail or web address
%
\maketitle
\index{\'Angel R. L\'opez-S\'anchez}
\index{C\'esar Esteban}
\index{Jorge Garc\'{\i}a-Rojas}
% Use the \index{} command to code your author index

\begin{abstract}
We present the main results of the PhD Thesis carried out by L\'opez-S\'anchez (2006) \cite{LS06}, in which a detailed morphological, photometrical %%@
and spectros\-copical analysis of a sample of 20 Wolf-Rayet (WR) galaxies was realized. The main aims are the study of the star formation and O and %%@
WR stellar populations in these galaxies and the role that interactions between low surface companion objects have in the triggering of the bursts. %%@
We analyze the morphology, stellar populations, physical conditions, chemical abundances and kinematics of the ionized gas, as well as the %%@
star-formation activity of each system.   

\end{abstract}

\section{Introduction}
\label{sec:1}
% Always give a unique label
% and use \ref{<label>} for cross-references
% and \cite{<label>} for bibliographic references
% use \sectionmark{}
% to alter or adjust the section heading in the running head
%Your text goes here. Use the \LaTeX\ automatism for your citations

WR galaxies are a subtype of H\,\textsc{ii}\ galaxies whose integrated spectra show broad emission lines attributed to WR stars, indicating the %%@
presence of an important population of massive stars and the youth of the starburst. Studying a sample of WR galaxies, \cite{Mendez99} and %%@
\cite{ME00} suggested that interactions with or between dwarf objects could be the main star formation triggering mechanism in dwarf galaxies and %%@
noted that the interacting and/or merging nature of WR galaxies can be detected only when deep and high-resolution images and spectra are available. %%@
Subsequent works (i.e., \cite{IPV01,VM01,VM02,Tran03}) also found a relation between massive star formation and the presence of interaction %%@
signatures in this kind of galaxies. Therefore, we have performed a detailed analysis of a sample of 20 of these objects extracted from the latest %%@
catalogue of WR galaxies \cite{SCP99} combining deep optical and near-infrared (NIR) broad band and H$\alpha$ ima\-ging together with optical %%@
spectroscopy (long slit and echelle) data. Additional X-ray, far-infrared and radio data were compiled from literature.

\begin{figure}[t]
\centering
\includegraphics[width=11.7cm]{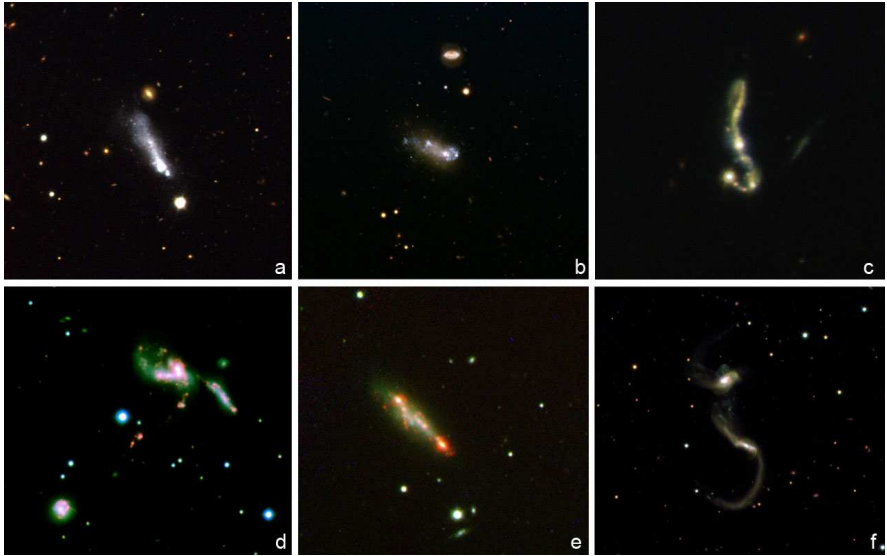}
\caption{\small{False color images of six of the WR galaxies analyzed. (a) SBS 1415+437 combining images in $B$ (blue), $V$ (green) and $R$ (red) %%@
filters. (b) SBS 1054+365 using $U$ + $B$ + $V$. (c) IRAS 08208+2816 using $U$ + $B$ + $V$. (d) HCG 31 using $V$ + $R$ + H$\alpha$. (e) SBS 1319+579 %%@
using $B$ + $R$ + H$\alpha$. (f) Arp 252 using $U$ + $B$ + $R$.}}
\label{fig:1}      
\end{figure}

\section{Optical and NIR imaging}
\label{sec:2}
Deep and high spatial resolution imagery in optical and NIR broad band filters have been used to study the morphology of the stellar component of the %%@
galaxies, looking for morphological features that reveal interaction processes with external gala\-xies or low surface brightness objects. Optical %%@
imagery was mainly obtained with ALFOSC instrument at 2.56m NOT, but also CAFOS at 2.2m CAHA and WFC and 2.5m INT were used. In Figure~\ref{fig:1} we %%@
show false color images of 6 galaxies of our sample. The quali\-ty of the data has allowed to detect faint features surrounding the galaxies, %%@
including tails (i.e, IRAS 08208+2816 and Arp 252, see Figure~\ref{fig:1}), independent dwarf galaxies (i.e., Mkn 1087 \cite{LSER04b}), mergers (i.e. %%@
HCG 31 \cite{LSER04a}) and candidate to tidal dwarf galaxies (TDGs) and arcs (i.e., IRAS 08339+6517, \cite{LSEGR06}).

\begin{figure}[t]
\centering
\includegraphics[height=5.5cm]{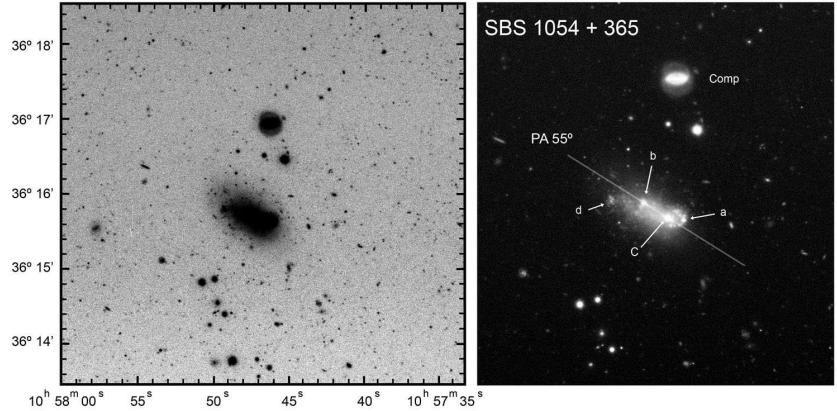}
\caption{\small{(\emph{Left}) Deep optical image of the WR galaxy SBS 1054+365 combining data in $U$, $B$ and $V$ filters. It has been saturated to %%@
reveal weakest features. (\emph{Right}) The same object in a non-satured image. Some important regions inside the galaxy and the position of the slit %%@
used for spectroscopy are indicated. The grayscale is in logarithmic scale in both cases.}}
\label{fig:1b}
\end{figure}

\begin{figure}[t]
\centering
\includegraphics[height=5.8cm]{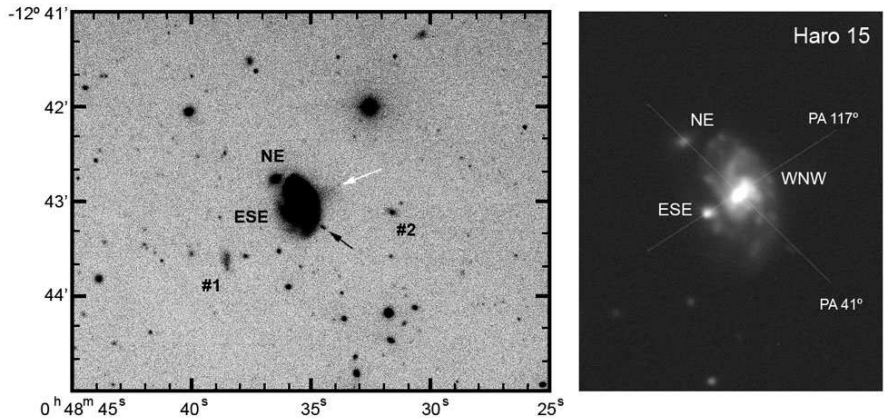}
\caption{\small{(\emph{Left}) Deep optical image of the WR galaxy Haro 15 in $R$ filter, that has been saturated to reveal weakest features. %%@
(\emph{Right}) The same object in a non-satured image. Some important regions inside the galaxy and the position of the slits used for spectroscopy %%@
are indicated. The grayscale is in logarithmic scale in both cases.}}
\label{fig:1c}     
\end{figure}

Figures~\ref{fig:1b}, \ref{fig:1c}, \ref{fig:2b}a and \ref{fig:3c} show some examples of our morphological analysis: the blue compact dwarf galaxy %%@
(BCDG) SBS 1054+365 (Figure~\ref{fig:1b}), the blue compact galaxy (BCG) Haro 15 \cite{LSE03} (Figure~\ref{fig:1c}; this image reveals that Haro 15 %%@
is in interac\-tion with two nearby objects labeled as ESE and NE), the infrared luminous galaxy IRAS 08208+2818 (Figure~\ref{fig:2b}) and Tol 9 %%@
(Figure~\ref{fig:3c}). The photometric analysis of the galaxies and the use of population synthesis mo\-dels (i.e. STARBURTS99 \cite{L99}; PEGASE.2, %%@
\cite{PEGASE97}) has permitted to analyze their colors, stellar populations (young, intermediate and old) and the age of the last star-forming burst.

An interesting object is the WR galaxy Tol 9 within the Klemola 13 group, located a 43.3 Mpc. Our false color image of Tol 9 is shown in %%@
Figure~\ref{fig:2}: several independent objects are found in its neighbourhood, being the more important the nearby spiral galaxy ESO 436-46 (at 96" %%@
$\sim$ 20.2 kpc). The images also reveal a bridge from Tol 9 towards a dwarf companion object located 10 kpc at SW, indicating probable interaction %%@
phenomena between both galaxies. No ionized gas is detected in the bridge, indeed, the analysis of its optical and NIR colors suggests that it is %%@
mainly dominated by a relatively old stellar population with ages higher than 500 Myr.

\section{H$\alpha$ imaging}
\label{sec:3}

\begin{figure}[t]
\centering
\includegraphics[width=11.7cm]{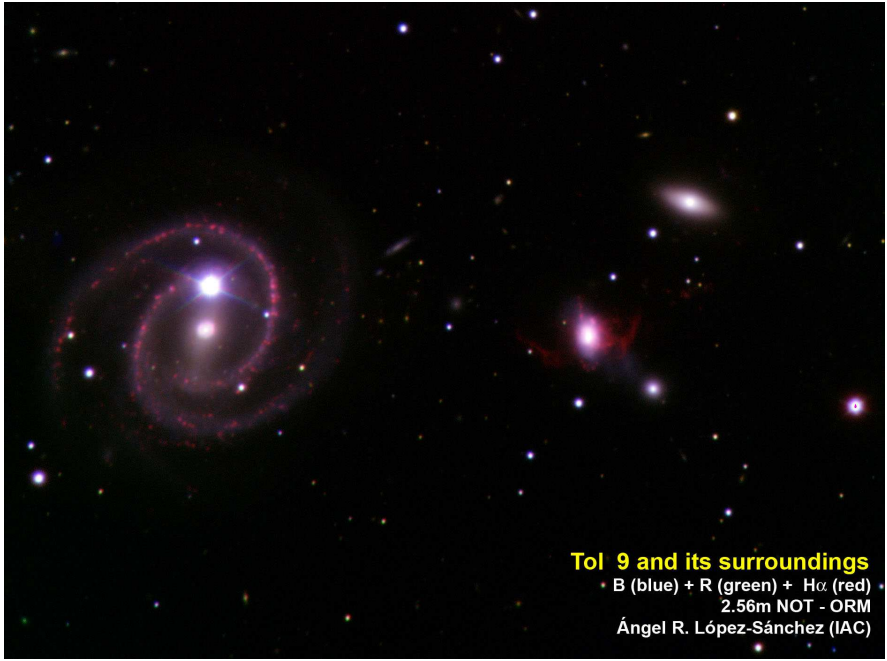}
\caption{\small{False color image of starburst galaxy Tol 9 (left) and the beauty spiral ESO 436-46 (left) within the Klemola 13 group, combining %%@
data in $B$ (blue), $R$ (green) and H$\alpha$ filters obtaining using ALFOSC at 2.56m NOT. Note the peculiar H$\alpha$ distribution of Tol 9, %%@
suggesting a kind of galactic wind in the galaxy.}}
\label{fig:2}      
\end{figure}

Deep continuum-subtracted H$\alpha$ images have been used to know the distribution and intensity of the ionized gas throughout the galaxies, see %%@
Figure~\ref{fig:1} d (HCG 31) \& e (SBS 1319+579), Figure~\ref{fig:2} (Tol 9), Figure~\ref{fig:2b}b (IRAS 08208+2816) and Figure~\ref{fig:3c}c (Tol %%@
9) as examples.  The data have been also used to estimate the H$\alpha$ luminosity, the number of ionizing stars, the mass of the ionized gas and the %%@
star formation rate (SFR) of each burst. The SFR derived from H$\alpha$ data is compared with that obtained using other methods. 

As we see in Figures~\ref{fig:2} and \ref{fig:3c}c, the continuum-subtracted H$\alpha$ emission map of Tol~9 reveals a kind of filamentary structure %%@
that is more extended than that seen in broad-band filters, suggesting that an outflow of material or a galactic wind exists in the starburst. The %%@
total H$\alpha$ luminosity of this galaxy indicates a total ionized mass of $M_{H\,II}$ = (3.4 $\pm$ 0.2) $\times$ 10$^6$ $M_{\odot}$ and a star %%@
formation rate of $SFR_{H\alpha}$ = 1.82 $\pm$ 0.13 $M_{\odot}$ a$^{-1}$.

\section{Intermediate resolution spectra}
\label{sec:4}

\begin{figure}[h]
\centering
\includegraphics[height=5.5cm]{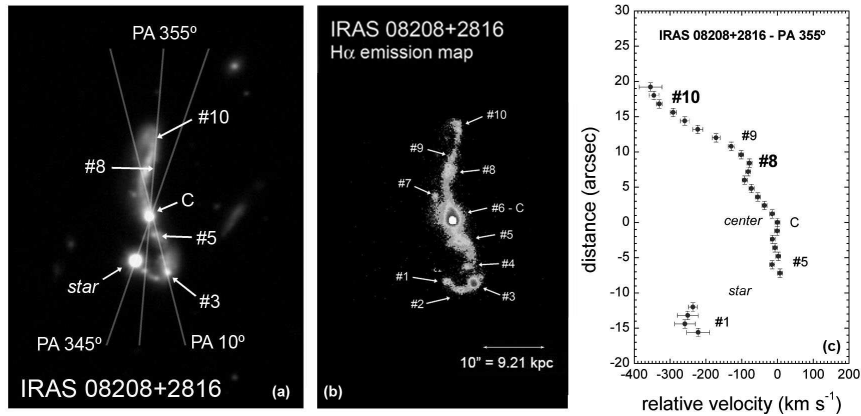}
\caption{\small{(a) Deep optical image of the WR galaxy IRAS 08208+2816 showing the position of the slits used for spectroscopy. (b) %%@
Continuum-subtracted H$\alpha$ of this galaxy labeling all the different objects. (c) Position-velocity diagram for the slit position with PA %%@
355$^{\circ}$, in which a tidal tail is clearly detected.}}
\label{fig:2b}       % Give a unique label
\end{figure}

Long slit and echelle (only for NGC 5253, see \cite{LSEGRPR07} and \S\ref{sec:7}) spectroscopy have been used to study the physical conditions %%@
(electron density and temperature, reddening, ionization nature), the chemistry of the gas (chemical abundances of He, O, N, S, Ne, Ar, Fe, Cl) and %%@
the kinematics of the ionized gas, as well as the massive star population content and its spatial location in each galaxy. The metallicity of each %%@
galaxy has be estimated by the direct method (assuming that electron temperature is known) and by the so-called empirical calibrations. 

\begin{figure}[t]
\centering
\includegraphics[angle=0,width=11.7cm]{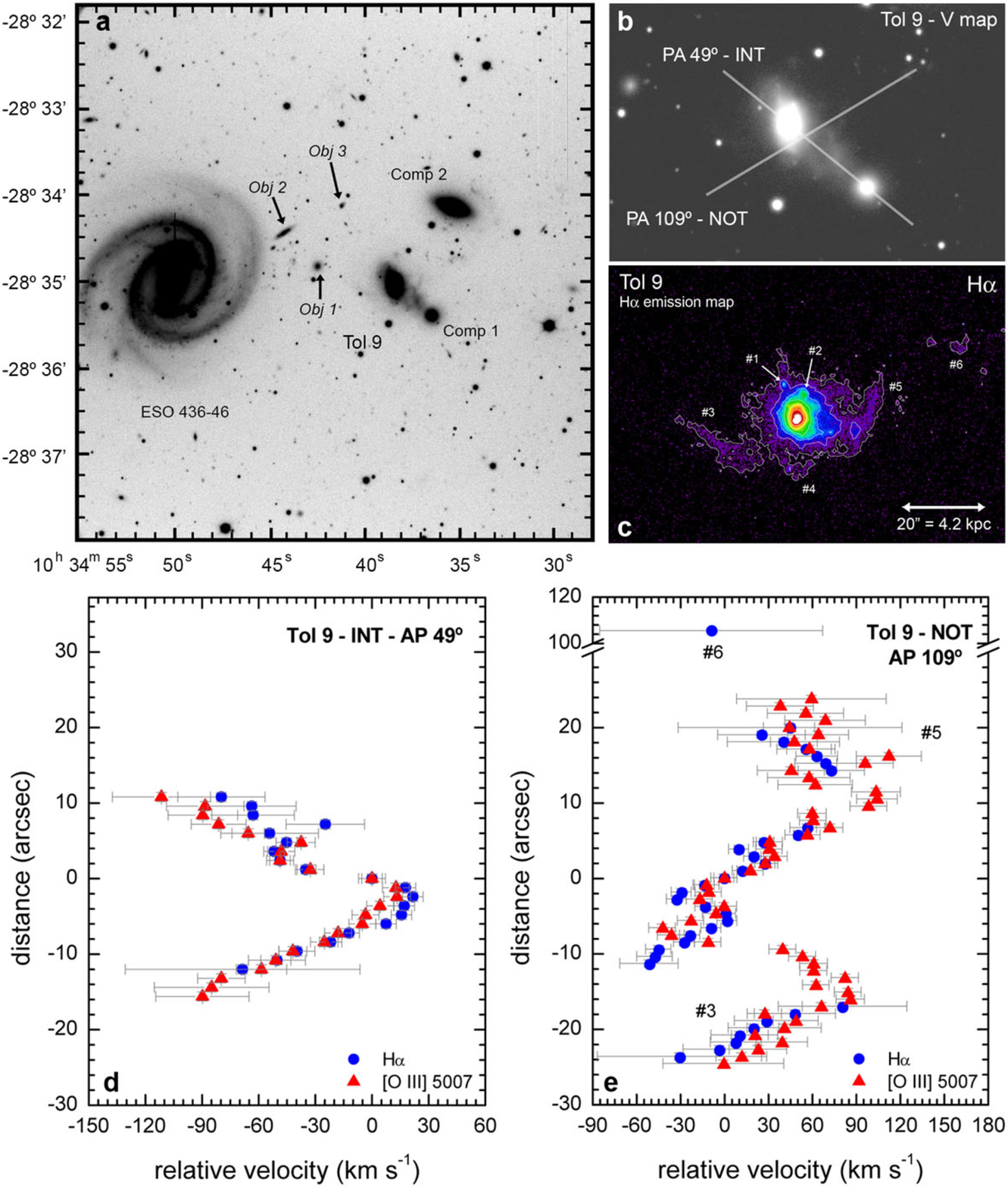}
\caption{\small{(a) Deep optical image of Tol 9 and its surroundings in $V$ filter (2.56m NOT). (b) Non-saturated $V$ image showing the slit %%@
positions used for spectroscopy. (c) Continuum-subtracted H$\alpha$ emission map of Tol 9. (d \& e) Position-velocity diagrams of Tol 9 obtained with %%@
our long-slit spectroscopy using 2.5m INT (AP 49$^{\circ}$) and 2.56m NOT (AP 109$^{\circ}$). The kinematics of the ionized gas was analyzed via the %%@
study of the profiles of the H$\alpha$ and [O III] $\lambda$5007 emission lines. North is up in both diagrams.}}
\label{fig:3c}       % Give a unique label
\end{figure}

\begin{figure}[t]
\centering
\includegraphics[angle=270,width=12cm]{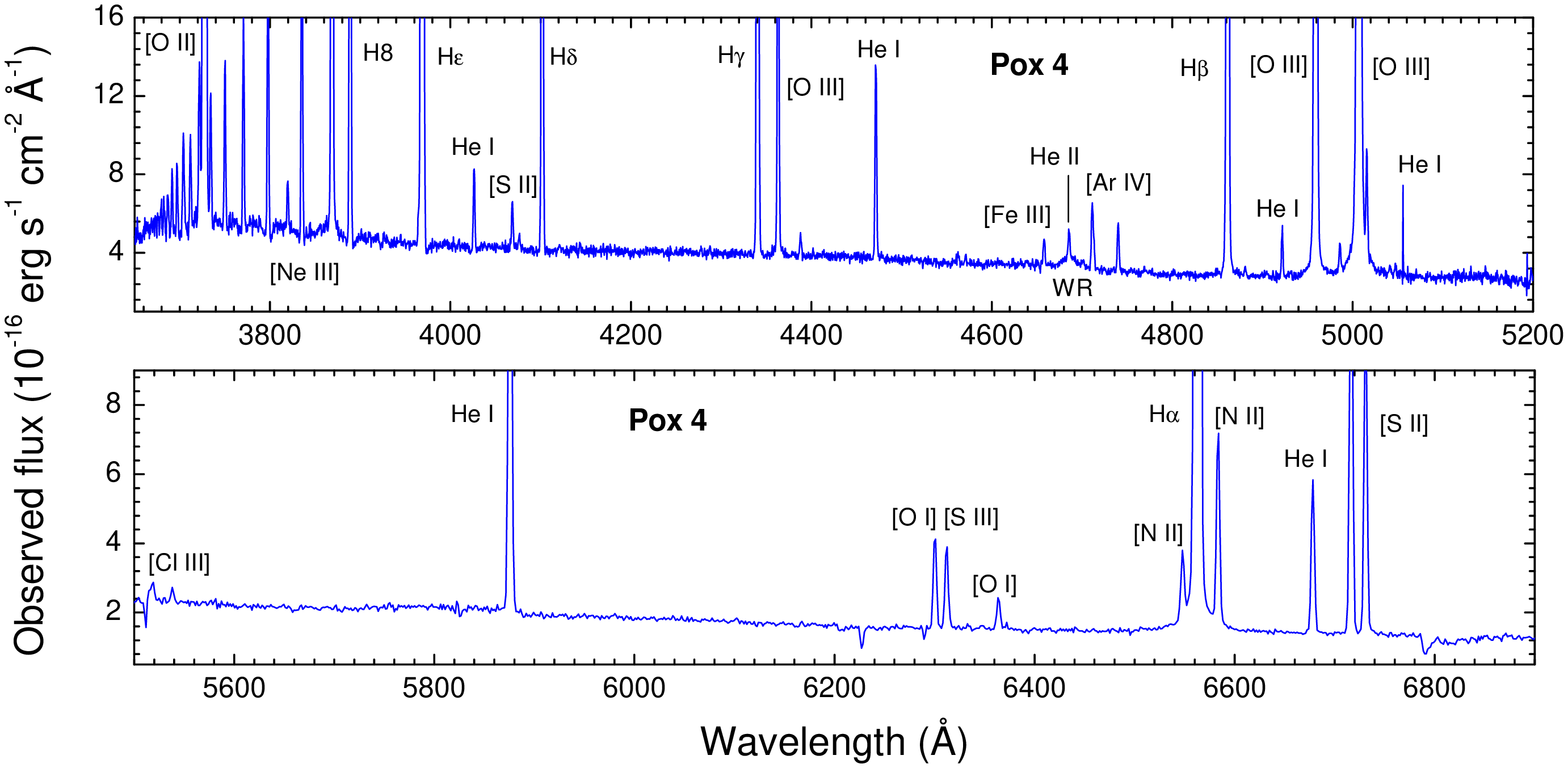}
\caption{\small{Redshift-corrected spectrum of POX 4 obtained with ISIS spectrograph at 4.2m WHT. The most important emission lines are labeled. Note %%@
the He~\textsc{ii} emission line at $\lambda$4686, indicating the presence of Wolf-Rayet stars.}}
\label{fig:3b}       % Give a unique label
\end{figure}

In Figure~\ref{fig:3b} we show our deep spectrum of the BCDG POX 4, showing the emission features that reveal the presence of WR stars (the so-called %%@
\emph{WR bump} and the He~\textsc{ii} emission line at $\lambda$4686). Our study also led to disentangle the tidal/pre-existing nature of the %%@
companion objects surrounding the main galaxies. For example, the oxygen abundance of object NE in Mkn 1199 is 12+log(O/H)=8.46, whereas its center %%@
has 8.75, confirming that NE is an external galaxy in interaction with it. This interaction triggers star-formation activity in the external zone of %%@
Mkn 1199. 

The kinematics of the ionized gas was studied via the analysis of emission line profiles using the 2D spectra. In objects in which solid-body %%@
rotation is found, the Keplerian mass have been estimated. Sometimes, prominent tidal tails (i.e. HCG~31 \cite{LSER04a}, Mkn~1087 \cite{LSER04b} and %%@
the evident case of IRAS~08208+2816, see Figure~\ref{fig:2b}) or outflows (Tol 9) have been detected. Indeed, the position-velocity diagram of Tol 9 %%@
using the NOT data (AP 109$^{\circ}$; see Figure~\ref{fig:3c}e) reveals a velocity pattern that can not be attributed to rotation; our analysis %%@
suggests that it could be explained considering a bipolar bubble expanding at about 80 km s$^{-1}$.

\section{Interactions and star formation in galaxy groups}
\label{sec:5}

Specially interesting are the cases of the groups of galaxies HCG 31 \cite{LSER04a} and Mkn 1087 \cite{LSER04b}, where interactions involving more %%@
than two objects are needed to explain the tails, bridges, mergers and tidal dwarf galaxies (TDGs) observed in them. We find that member F in HCG 31 %%@
hosts the youngest starburst of the group ($\sim$2.5 Myr), showing a substantial population of WR stars and, like member E, seems to be a TDG formed %%@
from material stripped from the brightest galaxy of the group. 

\begin{figure}[t]
\centering
\includegraphics[width=11.7cm]{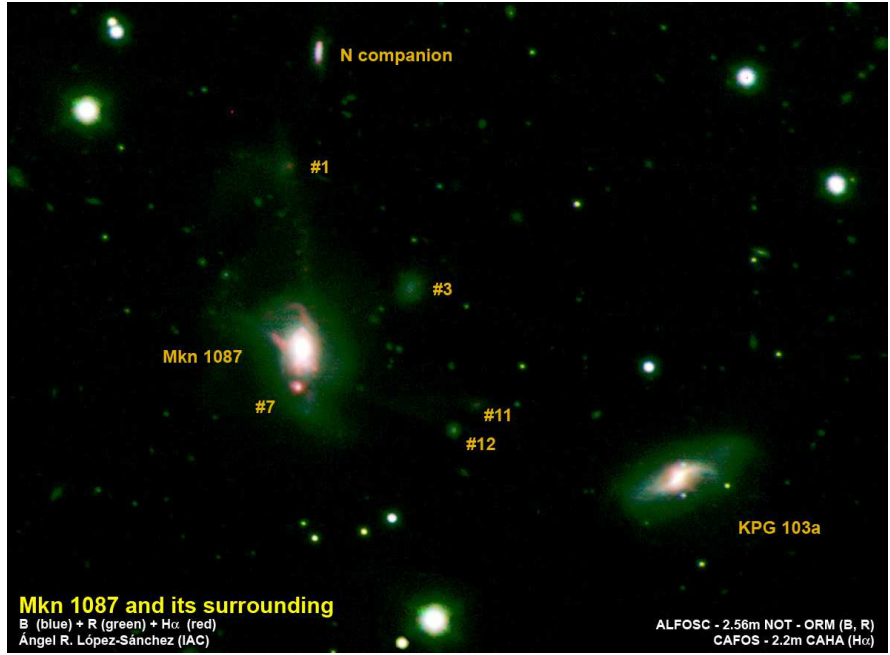}
\caption{\small{False color image of Mkn 1087 and its surrounding combining data in $B$ (blue), $R$ (green) and H$\alpha$ (red). Image in $R$ filter %%@
has been saturated to reveal the weak features found between several objects. Mkn 1087 itself, its north companion object and the bright galaxy KPG %%@
103 are the main members of a group of galaxies.}}
\label{fig:3d}      
\end{figure}

On the contrary, one nearby dwarf companion object located at the north of Mkn 1087 (see Figure~\ref{fig:3d}) is not a TDG but an independent dwarf %%@
galaxy that is in interaction with it. A nearby encounter between them created a long tidal tail. However, other surrounding objects seem to be TDGs %%@
(objects \#1, \#3, \#11 and \#12 in Figure~\ref{fig:3d}) Mkn 1087 can be classified as a low-z Luminous Compact Blue Galaxy (LCBG), rare objects in %%@
the local Universe but common at high redshift. LCBGs are especially interesting for studies of galaxies evolution and formation because they could %%@
be the equivalent of the high-z Lyman-break galaxies in the local universe \cite{EP03}.  

\section{Star formation and stellar populations \\ in IRAS 08339+6517}
\label{sec:6}

An excellent example of the analysis we have performed for each system is found in our detailed study of the star formation activity and stellar %%@
populations in the bright starburst galaxy IRAS 08339+6517 \cite{LSEGR06}. All data (broad and narrow-band imagery and spectra) were obtained using %%@
ALFOSC at NOT. Our new deep images reveal interactions features between IRAS 08339+6517 and a nearby dwarf object, as well as strong H$\alpha$ %%@
emission in its inner part. The chemical composition of the ionized gas of both galaxies is rather similar. Our deep spectra seem to show, for the %%@
first time, WR features in the center of IRAS 08339+6517. The kinematics also indicates interaction features and reveals an object that could be a %%@
candidate tidal dwarf galaxy or a remnant of an earlier merger. Our data suggest that a prominent H~\textsc{i} tidal tail found by \cite{CSK04} has %%@
been mainly formed from material stripped from the main galaxy.

\begin{figure}[t!]
\centering
\includegraphics[angle=90,width=8.3cm]{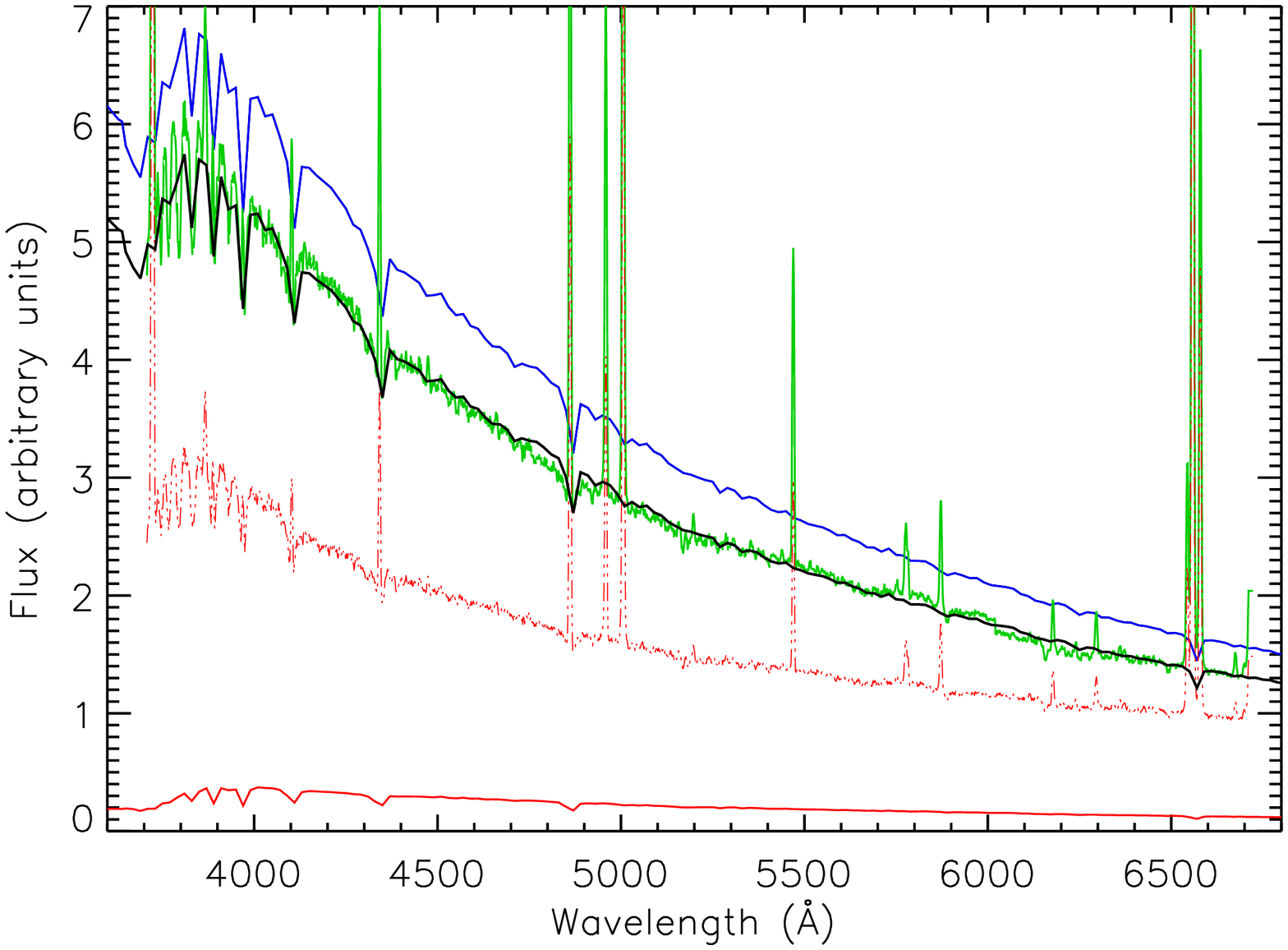}
\includegraphics[angle=90,width=8.3cm]{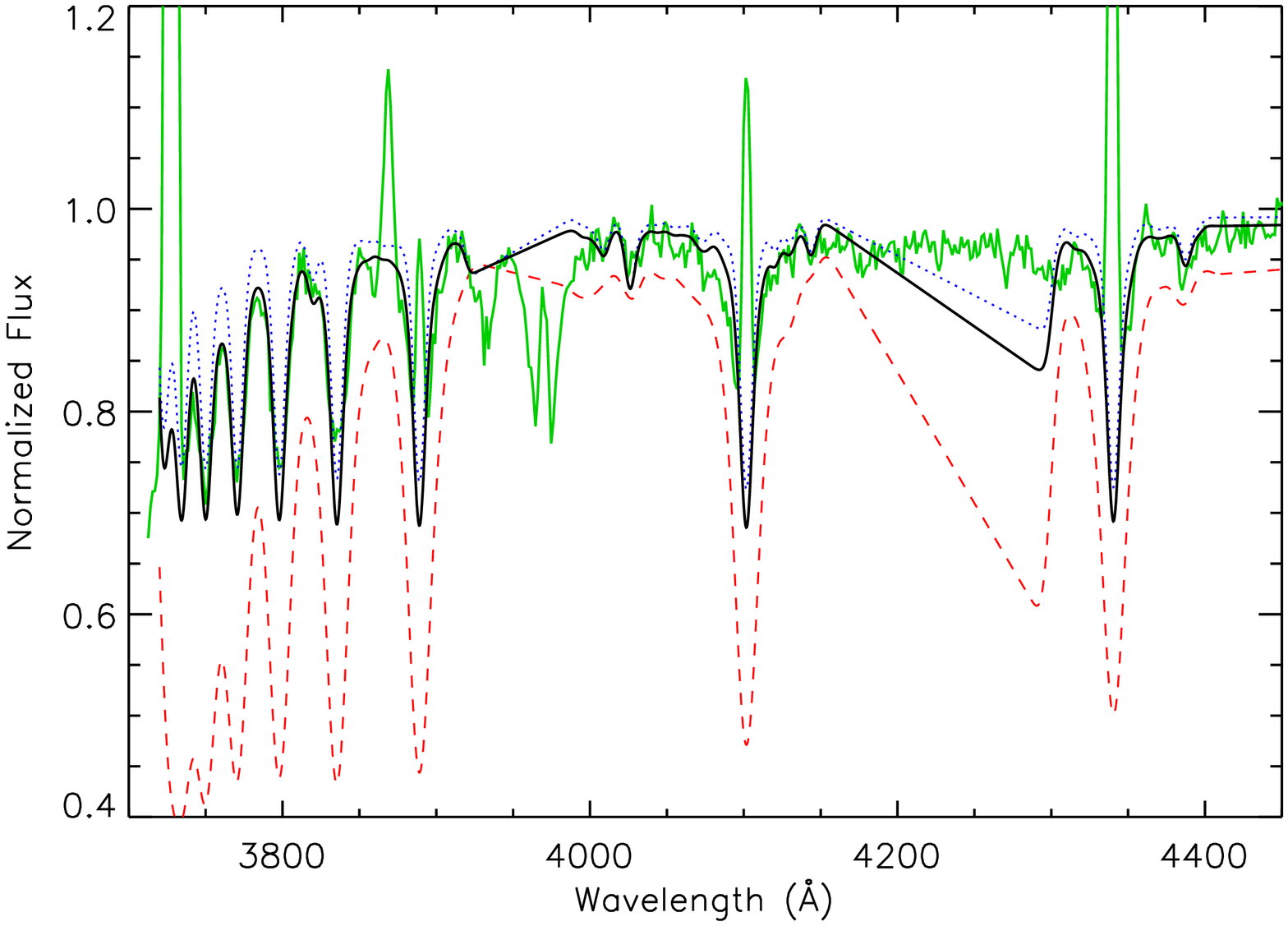}
\caption{\small{(\emph{Top}) Spectrum of IRAS 08339+6517 (obtained with ALFOSC at NOT) compared with synthetic continuum spectral energy %%@
distributions obtained using the PEGASE.2 \cite{PEGASE97} code. The dotted red line represents the observed spectrum uncorrected for extinction, %%@
whereas the green continuous line is the extinction-corrected spectrum. The upper conti\-nuous blue line corresponds to a model with an age of 6 Ma %%@
(young population model), whereas the lower continuous red line is a 140 Ma model (old population model). The shape of our observed derredened %%@
spectrum fits with a model with a contribution of 85\% for the young population and 15\% for the old population is considered (continuous black line %%@
over the galaxy spectrum). (\emph{Down}) Normalized dereddened spectrum of IRAS 08339+6517 (green continuous line) compared with \cite{GD99} models %%@
with 4 Ma (dotted blue line) and 200 Ma (dashed line) at $Z_{\odot}$ metallicity. The best fit corresponds to a model with a contribution of 85\% for %%@
the young population and 15\% for the old population (continuous black line). Note that the models only give values for the position of the %%@
absorption lines and not for all wavelengths, being the reason of the straight lines presented between 3930 and 4000 \AA\ and 4150 and 4300 \AA.}}
\label{fig:3}
\end{figure}

We estimate that the age of the most recent burst is around 4 -- 6 Ma, but a more evolved underlying stellar population, with a minimal age between %%@
100 -- 200 Ma, is also detected and fits an exponential intensity profile. A model which combines 85\% young and 15\% old populations can explain %%@
both the spectral energy distribution (see Figure~\ref{fig:3}, up) and the H~\textsc{i} Balmer and He~\textsc{i} absorption lines (see %%@
Figure~\ref{fig:3}, down) observed in our spectrum. We finally conclude that IRAS 08339+6517 does satisfy the criteria of Luminous Compact Blue %%@
Galaxy (LCBG), rare objects in the local Universe but common at high redshift.  There are very few local LCBGs nowadays detected but nearly half of %%@
them have optical companions, present disturbed morphologies and/or are clearly interacting \cite{GPW04}. If interactions were the responsible of the %%@
activity in LCBGs, it would indicate that they were perhaps more common at high redshifts, as the hierarchical galaxies formation models predict %%@
(i.e., \cite{KW93,Springel05}). This fact would also support the idea that interaction with dwarf companion objects could be an important trigger %%@
mechanism of the star formation activity in local starbursts.

\section{The localized chemical pollution in NGC~5253}
\label{sec:7}

\begin{figure}[t]
\centering
\includegraphics[angle=270,width=12cm]{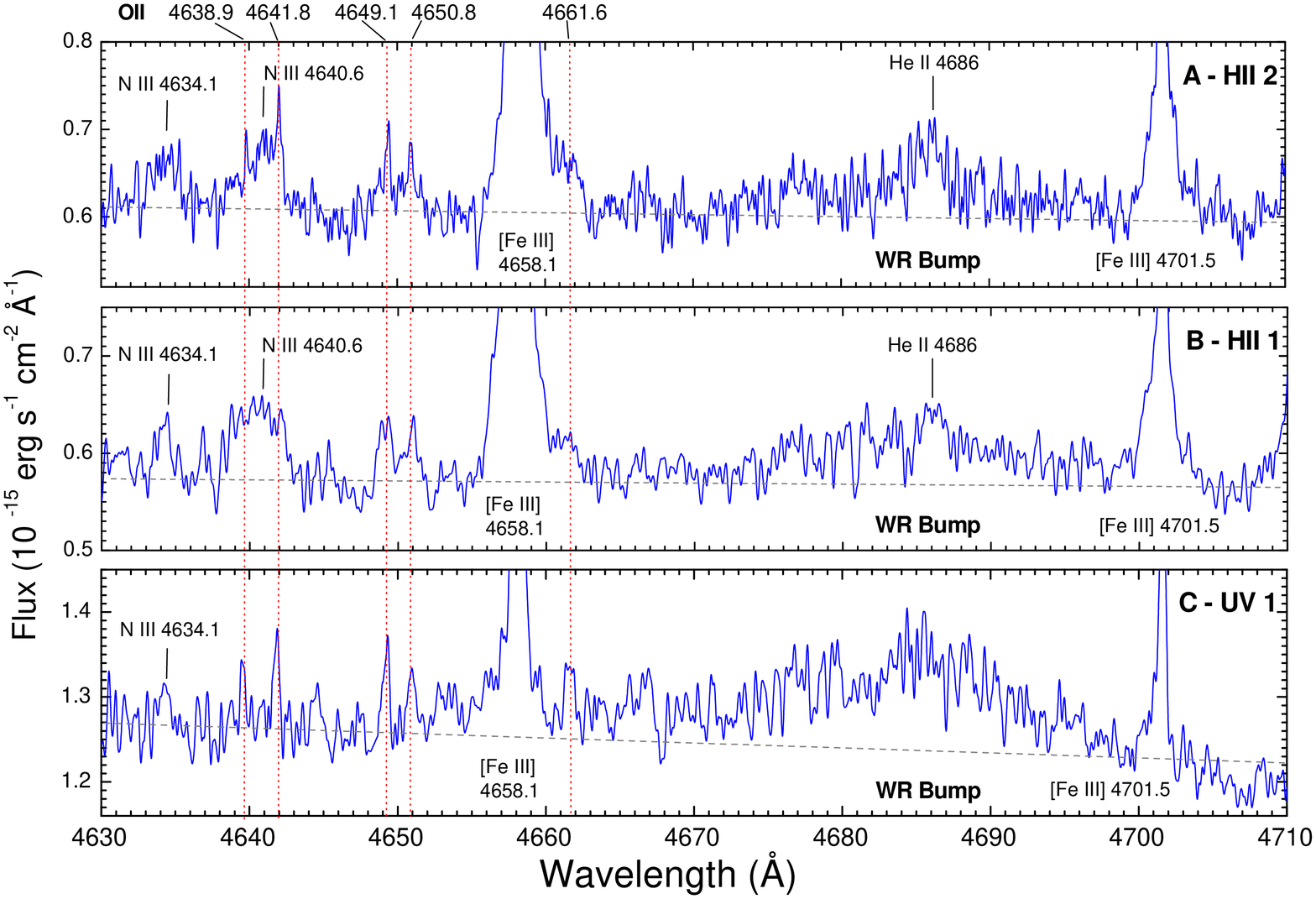}
\caption{\small{Sections of echelle spectra of zones A, B, and C of NGC 5253 showing the lines of  multiplet 1 of O~\textsc{ii} and the WR bump. Note %%@
the broad N~\textsc{iii} $\lambda$4641 emission line blended with O~\textsc{ii} $\lambda\lambda$4639,4642 emission lines in regions A and B. That %%@
line is absent in region C. A broad N~\textsc{iii} $\lambda$4634 seems to be also present in region A.}}
\label{fig:4b}       % Give a unique label
\end{figure}

One of our main goals is the detection of the weak O~\textsc{ii} and C~\textsc{ii} recombination lines in our deep VLT spectra of the dwarf galaxy %%@
NGC 5253, the first time reported in a starburst (see Figure~\ref{fig:4b} and \cite{LSEGRPR07}). The ionic abundances derived from the recombination %%@
lines are from 0.20 to 0.40 dex higher than those calculed from collision excited lines, in agreement with the result found in other Galactic and %%@
extragalactic H~\textsc{ii} regions. This conclusion suggests that temperature fluctuations may be present in the ionized gas of this galaxy. %%@
Furthermore, we detect a localized nitrogen enrichment in two of the central starburst of the galaxy, an well as a possible slight helium pollution %%@
in the same zones. The enrichment pattern completely agrees with that expected by the pollution of the winds of massive stars in the WR phase. The %%@
amount of enriched material needed to produce the observed overabundance is consistent with the mass lost by the number of WR stars estimated in the %%@
starbursts. 

\section{Global properties}
\label{sec:8}

\begin{figure}[t]
\centering
\includegraphics[angle=90,width=12cm]{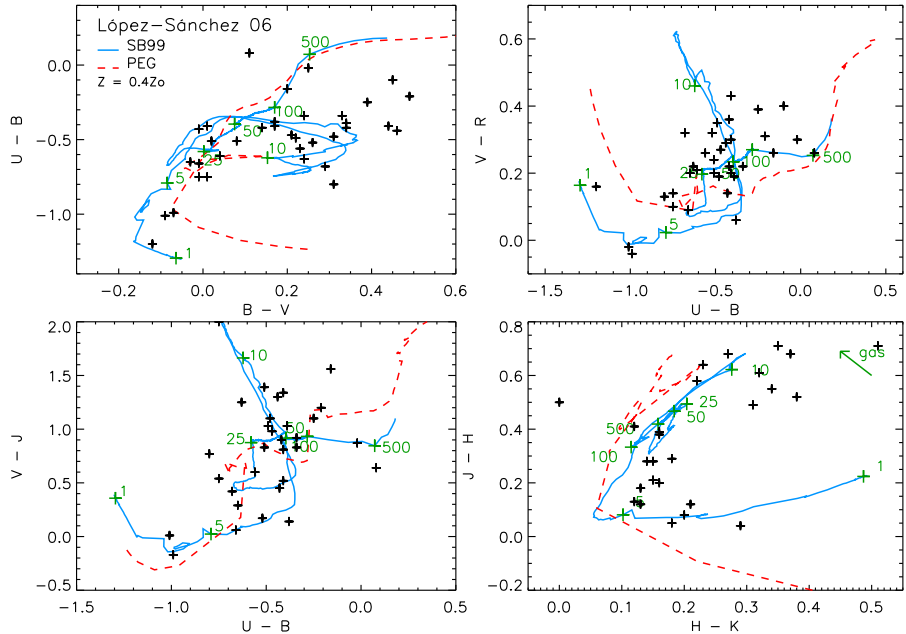}
\caption{\small{Color-color diagrams comparing the STARBURST99 (\cite{L99}; blue continuous line) and PEGASE.2 (\cite{PEGASE97}; red discontinuous %%@
line) theoretical models with the colors observed in the objects of our WR galaxies sample.  Some ages (in Ma) are indicated for STARBURST99 %%@
models.}}
\label{fig:5}       % Give a unique label
\end{figure}

Finally, we have performed a global analysis of our sample of 20 Wolf-Rayet galaxies. It is the more complete and exhaustive data set of this kind of %%@
galaxies, involving multiwavelength results and being every one analyzed following the same procedures. The main global results are the following:

\begin{enumerate}
\item The analysis of WR features in our sample suggests that aperture effects and localization of the bursts with WR stars seem to play a %%@
fundamental role in the detection of this sort of massive stars in starburst galaxies. 

\item Photometric data have been corrected for both extinction and nebular emission using our spectroscopic values. As we see in Figure~\ref{fig:5}, %%@
a good agreement between our optical and NIR colors and theoretical models is found; small discrepancies are explained by the existence of several %%@
(young and older) stellar populations.

\begin{figure}[t!]
\centering
\includegraphics[angle=270,width=7.6cm]{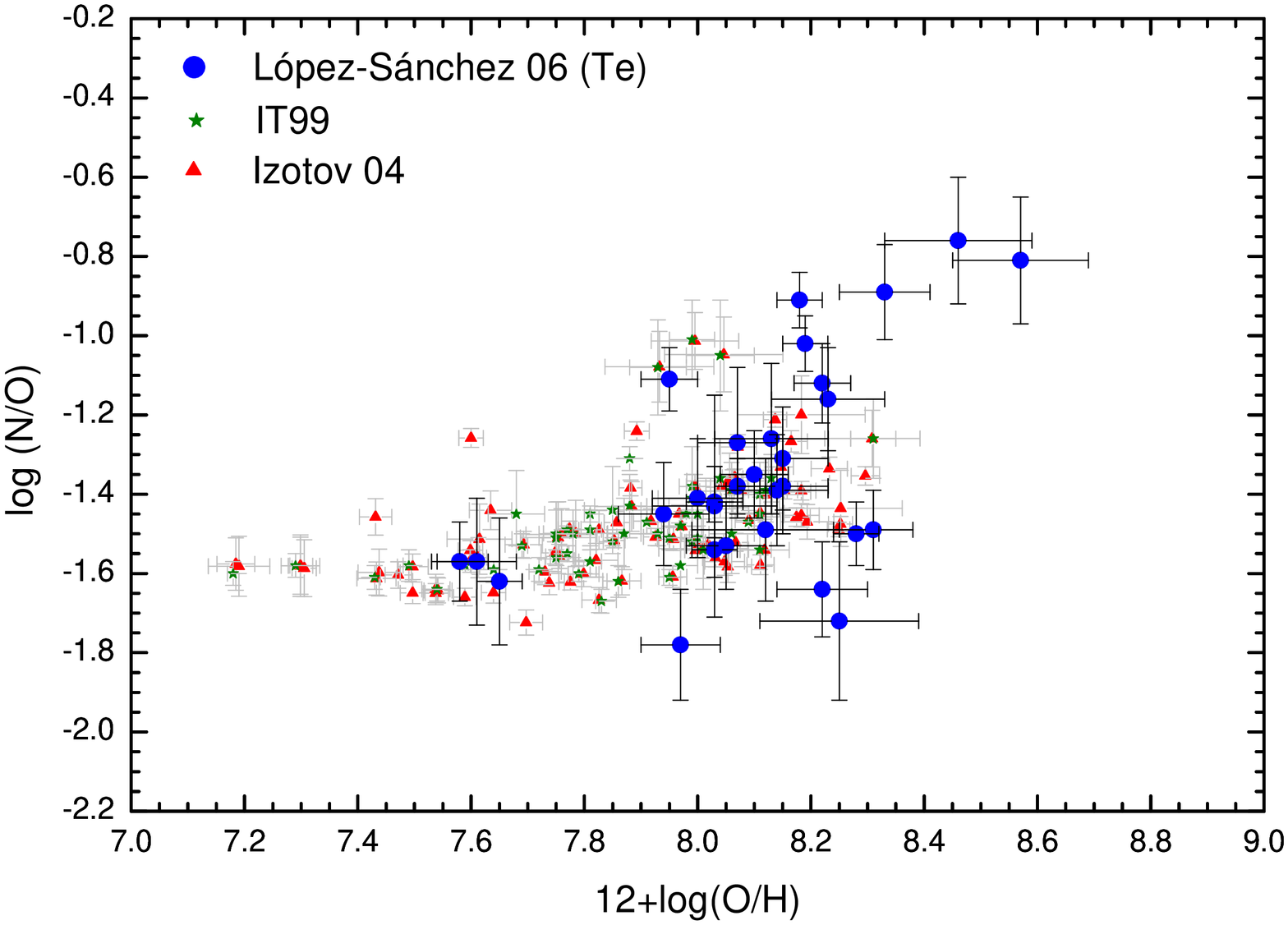}
\includegraphics[angle=270,width=7.6cm]{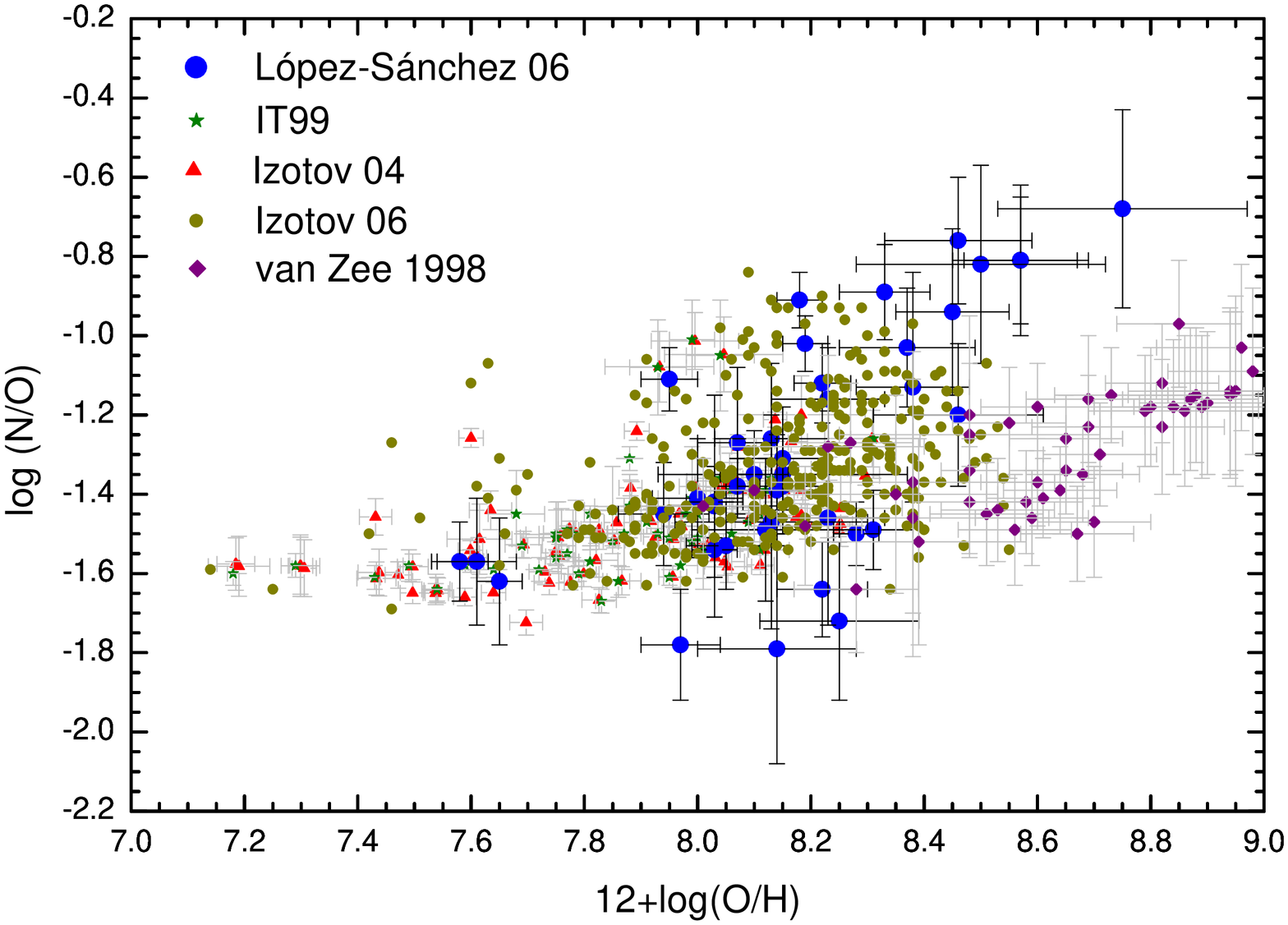}
\caption{\small{N/O vs. O/H diagram comparing the data obtained for our sample with previous analysis of BCDGs. (\emph{Top}) Blue circles: Data with %%@
direct estimation of electron temperature. Green stars: Data presented by Izotov \& Thuan (1999) \cite{IT99}. Red triangles: Izotov et al. (2004) %%@
\cite{IPG04} data. (\emph{Down}) Blue circles: all data obtained for our galaxy sample using both direct and empirical methods. Yellow stars: Izotov %%@
et al. (2006) \cite{ISMGT06} data. Purple rhombuses: van Zee et al. (1998) \cite{vZee98} data. Green stars and red triangles as the same that %%@
before.}}
\label{compabun}
\end{figure}

\item Physical and chemical properties are in agreement with both previous observations and models of chemical evolution of galaxies (see %%@
Figure~\ref{compabun}). We have compared abundances obtained by the direct method with those obtained for several empirical calibrations: Pilyugin %%@
(2001a,b) \cite{P01a,P01b} seems to give similar results whereas calibrations based in photoionization mo\-dels such as McGaugh (1994) \cite{M94} and %%@
Kewley \& Dopita (2002) \cite{KD02} give abundances higher ($\sim$0.20 dex) than expected.

\begin{figure}[t!]
\centering
\includegraphics[angle=270,width=\linewidth]{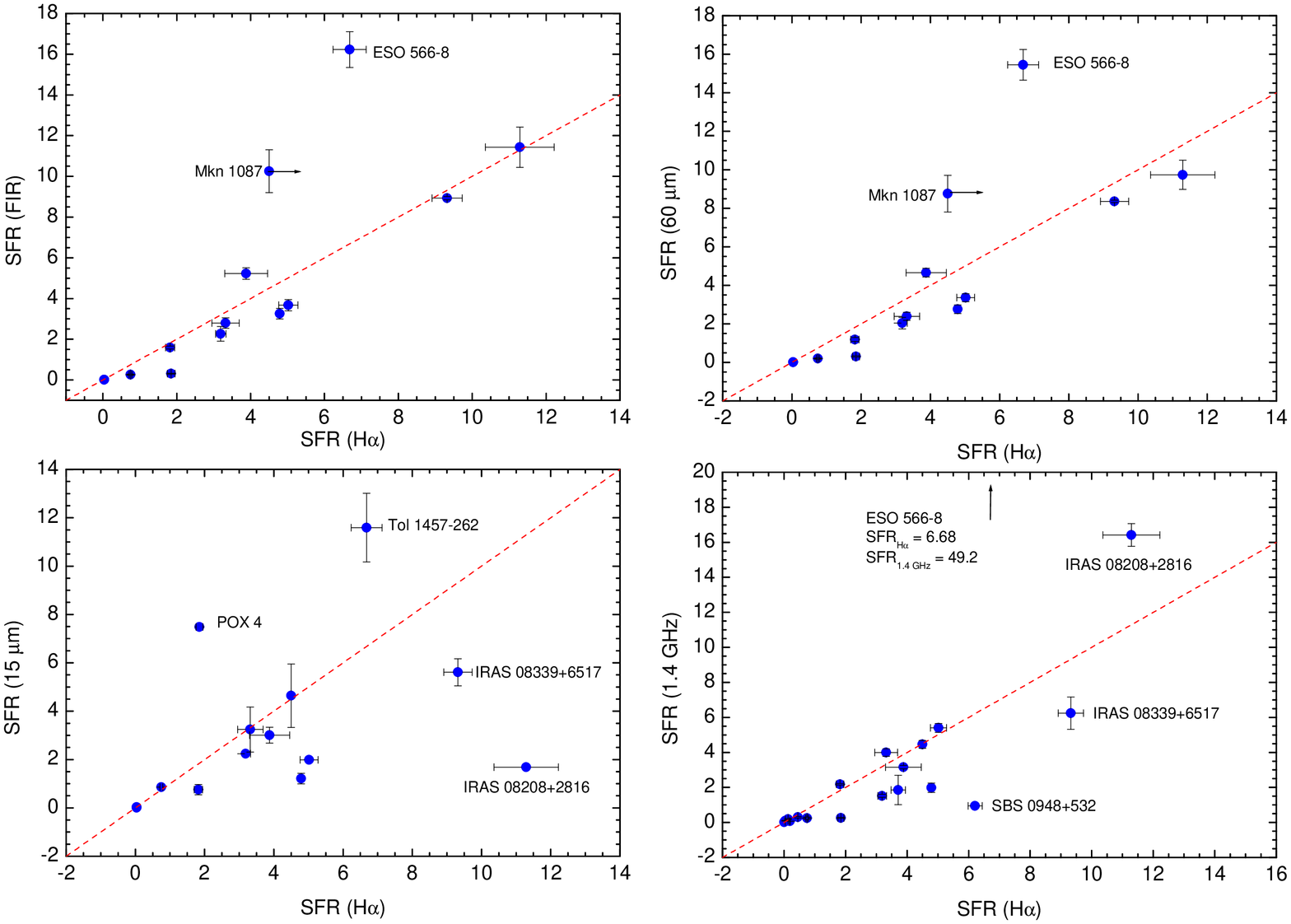}
\protect\caption[ ]{\small{Comparison between the SFR  determined using our H$\alpha$ flux (corrected by both extinction and [N II] emission, %%@
abscissa axis) with those obtained using far-infrared (FIR), 15 $\mu$m, 60 $\mu$m and 1.4 GHz radio-continuum luminosities. The H$\alpha$ flux for %%@
Mkn 1087 is a lower limit. The galaxy ESO 566-8 seems to have some kind of radio-activity. Dotted lines indicate equal values for the SFR.}}
\label{sfr}
\end{figure}

\item The comparison of the SFR derived from our H$\alpha$ data (corrected by \emph{both} extinction and [N II] emission using our spectroscopic %%@
data) is in good agreement with the SFR obtained using multiwavelength relations (see Figure~\ref{sfr}). We have also derived an X-ray based SFR for %%@
this kind of starburst galaxies.

\begin{figure}[t!]
\includegraphics[angle=270,width=\linewidth]{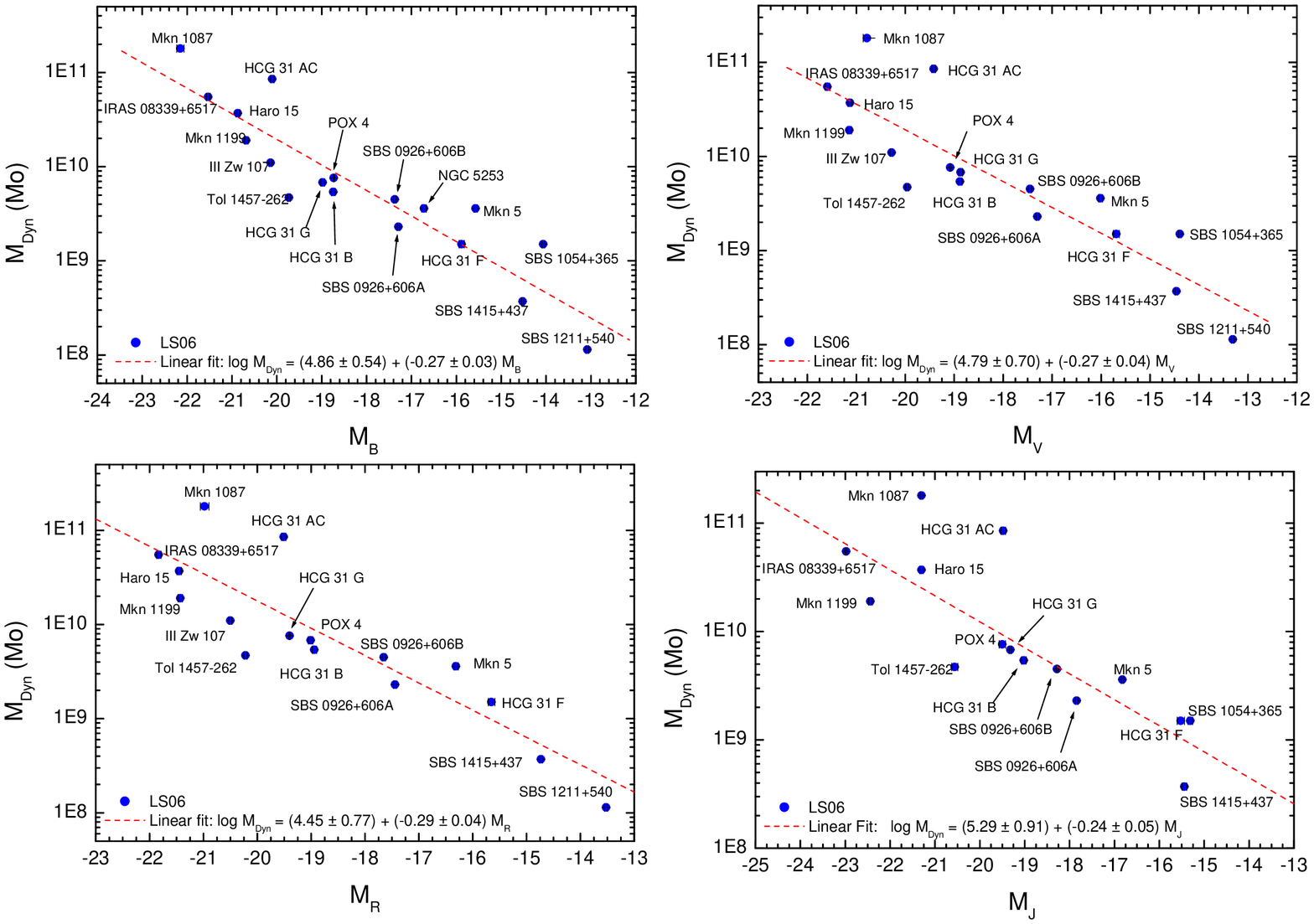}
\protect\caption[ ]{\small{Dynamical mass $M_{dyn}$ vs. absolute magnitude in  $B$, $V$, $R$ and $J$ for the individual galaxies analyzed in this %%@
work. A linear fit is also shown in each case.}}
\label{masascolor}
\end{figure}

\item We have determine the ionized gas mass ($M_{H\,II}$, using our H$\alpha$ images), neutral gas mass ($M_{H\,I}$, using 21 cm H I data from the %%@
literature), mass of the ionizing stellar cluster ($M_{\star}$), warm dust mass ($M_{dust}$, using FIR fluxes), Keplerian mass ($M_{kep}$, using the %%@
kinematics of the ionized gas) and the dynamical mass ($M_{dyn}$, considering the kinematics of the neutral gas). As it expected, all mass values %%@
increase with the luminosity of the galaxy (Figure~\ref{masascolor}). Furthermore, we find a good correlation between $M_{dyn}$ and the luminosity %%@
(absolute magnitude in $B$, $V$, $R$ and $J$ filters) of the galaxy. We also find a good relation between the reddening coefficient derived from the %%@
Balmer decrement, $C$(H$\beta$), and $M_{dust}$, as we see in Figure~\ref{mpolvo}. This fact suggests that extinction is mainly produced within the %%@
starburst.  

\begin{figure}[t!]
\centering
\includegraphics[angle=270,width=9.6cm]{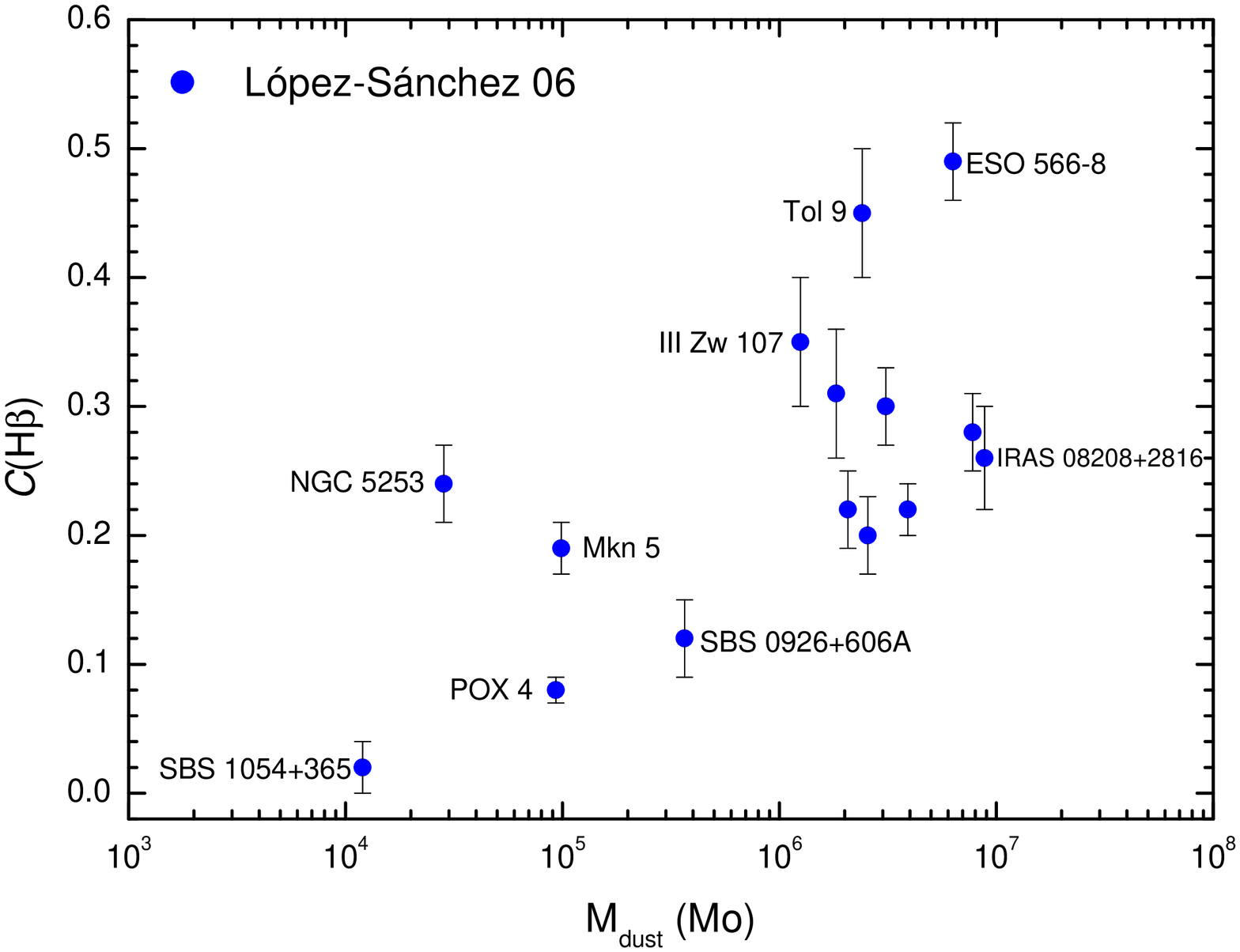}
\protect\caption[ ]{\footnotesize{Reddening coefficient, $C$(H$\beta$), vs. warm dust mass, $M_{dust}$, for all our galaxies for which both %%@
quantities are available.}}
\label{mpolvo}
\end{figure}

\end{enumerate}

\section{Conclusions}
\label{sec:9}

Our multiwavelength study has allowed to achieve a global vision of the star formation activity and evolution of each system, but also have permitted %%@
to find gene\-ral results involving all the galaxy sample. The main conclusion is that {\bf the majority of studied galaxies} (16 up to 20, %%@
$\sim$80\% of the objects) {\bf show clear interaction features} such as plumes, tails, TDGs, regions with very different chemical abundances inside %%@
galaxies, perturbed kinematics of the ionized gas or lack of neutral hydrogen gas, {\bf confirming the hypothesis that interaction with or between %%@
dwarf objects triggers the star formation activity in Wolf-Rayet galaxies}. 

\section{Acknowledgements}
\label{sec:10}

{\small \'A.R. L-S thanks C\'esar Esteban (his PhD supervisor) for all the help and very valuable explanations, talks and discussions during all last %%@
5 years. He also acknow\-ledges Jorge Garc\'{\i}a-Rojas and Sergio Sim\'on-D\'{\i}az for their help and friendship. He extends this acknowledge to %%@
all people at Instituto de Astrof\'{\i}sica de Canarias (IAC, Spain). \'A.R. L-S. {\bf deeply} thanks to Universidad de La Laguna (Spain) for force %%@
him to translate the PhD thesis from English to Spanish; nowadays he is translating from Spanish to English in order to publish his research.}

\printindex
\end{document}